\documentclass[floatfix,aps,prl,twocolumn, superscriptaddress]{revtex4-2}
\usepackage{upgreek}
\usepackage{graphicx}
\usepackage[]{epsfig}
\usepackage{times,amsmath,amssymb}

\usepackage{color}
\usepackage[colorlinks = true,
            linkcolor = blue,
            urlcolor  = blue,
            citecolor = blue,
            anchorcolor = blue]{hyperref}

\begin{document}

\title{Perfect impedance matching unlocks sensitive radio-frequency reflectometry in 2D material quantum dots}

\author{Motoya Shinozaki}
\email[]{motoya.shinozaki.c1@tohoku.ac.jp}
\affiliation{WPI Advanced Institute for Materials Research, Tohoku University, 2-1-1 Katahira, Aoba-ku, Sendai 980-8577, Japan}
\affiliation{Research Center for Materials Nanoarchitechtonics (MANA), National Institute for Material Science (NIMS),
1-2-1 Sengen, Tsukuba 305-0047, Japan}

\author{Akitomi Shirachi}
\affiliation{Research Center for Materials Nanoarchitechtonics (MANA), National Institute for Material Science (NIMS),
1-2-1 Sengen, Tsukuba 305-0047, Japan}
\affiliation{Research Institute of Electrical Communication, Tohoku University, 2-1-1 Katahira, Aoba-ku, Sendai 980-8577, Japan}
\affiliation{Department of Electronic Engineering, Graduate School of Engineering, Tohoku University, Aoba 6-6-05, Aramaki, Aoba-Ku, Sendai 980-8579, Japan}

\author{Yuta Kera}
\affiliation{Research Institute of Electrical Communication, Tohoku University, 2-1-1 Katahira, Aoba-ku, Sendai 980-8577, Japan}
\affiliation{Department of Electronic Engineering, Graduate School of Engineering, Tohoku University, Aoba 6-6-05, Aramaki, Aoba-Ku, Sendai 980-8579, Japan}

\author{Tomoya Johmen}
\affiliation{Research Institute of Electrical Communication, Tohoku University, 2-1-1 Katahira, Aoba-ku, Sendai 980-8577, Japan}
\affiliation{Department of Electronic Engineering, Graduate School of Engineering, Tohoku University, Aoba 6-6-05, Aramaki, Aoba-Ku, Sendai 980-8579, Japan}

\author{Shunsuke Yashima}
\affiliation{Research Institute of Electrical Communication, Tohoku University, 2-1-1 Katahira, Aoba-ku, Sendai 980-8577, Japan}
\affiliation{Department of Electronic Engineering, Graduate School of Engineering, Tohoku University, Aoba 6-6-05, Aramaki, Aoba-Ku, Sendai 980-8579, Japan}

\author{Aruto Hosaka}
\affiliation{Information Technology R\&D Center, Mitsubishi Electric Corporation, Kamakura 247-8501, Japan}

\author{Tsuyoshi Yoshida}
\affiliation{Information Technology R\&D Center, Mitsubishi Electric Corporation, Kamakura 247-8501, Japan}

\author{Takeshi Kumasaka}
\affiliation{Research Institute of Electrical Communication, Tohoku University, 2-1-1 Katahira, Aoba-ku, Sendai 980-8577, Japan}

\author{Yusuke Kozuka}
\affiliation{WPI Advanced Institute for Materials Research, Tohoku University, 2-1-1 Katahira, Aoba-ku, Sendai 980-8577, Japan}
\affiliation{Research Center for Materials Nanoarchitechtonics (MANA), National Institute for Material Science (NIMS),
1-2-1 Sengen, Tsukuba 305-0047, Japan}

\author{Tomohiro Otsuka}
\email[]{tomohiro.otsuka@tohoku.ac.jp}
\affiliation{WPI Advanced Institute for Materials Research, Tohoku University, 2-1-1 Katahira, Aoba-ku, Sendai 980-8577, Japan}
\affiliation{Research Institute of Electrical Communication, Tohoku University, 2-1-1 Katahira, Aoba-ku, Sendai 980-8577, Japan}
\affiliation{Department of Electronic Engineering, Graduate School of Engineering, Tohoku University, Aoba 6-6-05, Aramaki, Aoba-Ku, Sendai 980-8579, Japan}
\affiliation{Center for Science and Innovation in Spintronics, Tohoku University, 2-1-1 Katahira, Aoba-ku, Sendai 980-8577, Japan}
\affiliation{Research Center for Materials Nanoarchitechtonics (MANA), National Institute for Material Science (NIMS),
1-2-1 Sengen, Tsukuba 305-0047, Japan}
\affiliation{Center for Emergent Matter Science, RIKEN, 2-1 Hirosawa, Wako, Saitama 351-0198, Japan}

\date{\today}

\begin{abstract}
Two-dimensional (2D) materials are attractive platforms for realizing high-performance quantum bits (qubits).
However, radio-frequency (RF) charge detection, which is a key technique for qubits readout, remains challenging in such systems.
We demonstrate RF reflectometry with impedance matching for high-resistance quantum dot devices based on bilayer graphene and molybdenum disulfide. 
By integrating a tunable strontium titanate (SrTiO$_3$) varactor into a resonant circuit, we achieve nearly perfect impedance matching, enabling sensitive charge detection.
The demodulated RF signal clearly shows Coulomb oscillations, and the SrTiO$_3$ varactor exhibits robustness against both magnetic fields and voltage noise on the varactor. 
Our results establish SrTiO$_3$ varactors as effective tunable matching components for RF reflectometry in high-resistance 2D material quantum devices, providing a foundation for high-speed qubits readout.
\end{abstract}

\maketitle

\begin{figure}
\begin{center}
  \includegraphics{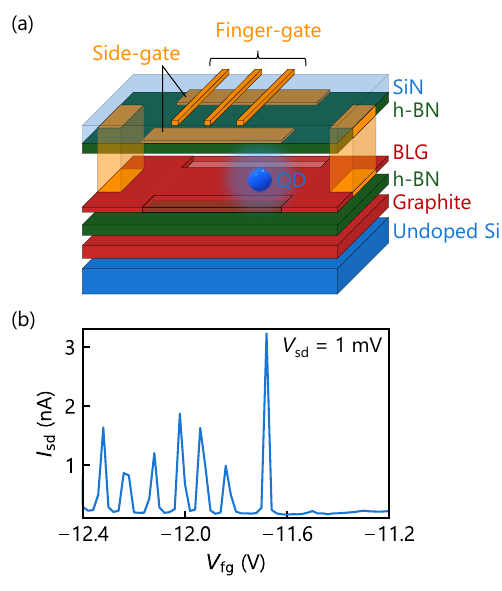}
  \caption{(a) Schematic of the device structure. A quantum dot (QD) is defined electrostatically in a narrow channel beneath the finger-gate electrode.  
  (b) Source–drain current $I_\mathrm{sd}$ as a function of finger-gate voltage $V_\mathrm{fg}$ at $V_\mathrm{sd}=1$~mV.}
  \label{fig1}
\end{center}
\end{figure}

\section{I. Introduction}

Two-dimensional (2D) materials have attracted much attention as promising platforms for quantum information processing, offering unique quantum bits (qubits) utilizing spin and valley states~\cite{xiao2012coupled, novoselov20162d, Eich2018spin, banszerus2022spin, garreis2024long, Shandilya2025}.
In particular, gate-defined quantum dots in bilayer graphene (BLG)~\cite{banszerus2018gate} and transition-metal dichalcogenides such as molybdenum disulfide (MoS$_2$)~\cite{lee2016coulomb, kumar2023excited, tataka2024surface} provide an attractive combination of clean electrostatic control, low nuclear-spin environments, and valley degrees of freedom.
Among these, the Kramers qubit protected by spin- and valley- mixing mechanisms has been experimentally demonstrated in carbon-nanotubes~\cite{laird2013valley} and gate-defined BLG quantum dots~\cite{banszerus2021spin, banszerus2022spin, banszerus2023particle} and shown to exhibit long coherence times~\cite{Denisov2025}.
Beyond this, Kramers qubits are expected to allow high-fidelity quantum operations and require a large number of rapid single-shot measurements for accurate characterization of gate control errors~\cite{Yoneda2014fast, 2018YonedaNatureNano}.
Most of the present measurements are still limited to direct current transport in gate-defined 2D material quantum dots, leading to long laboratory times to perform error-rate measurements.
Therefore, it is important to refine high-speed readout techniques for evaluating qubit performance in 2D materials.

Radio-frequency (RF) reflectometry plays an important role in fast and sensitive charge state probing in semiconductor quantum dots~\cite{vigneau2023probing}. 
Two complementary readout modes are typically employed. 
One is a dispersive readout, in which gate electrodes sense capacitance changes~\cite{Colles2013dispersive, west2019gate}.
The other is a resistive readout we focus here, monitoring source–drain conductance variations~\cite{reilly2007fast, barthel2009rapid}.
The dispersive readout is advantageous for scalability, whereas the resistive readout provides high sensitivity when combined with charge-sensing quantum dots. 
While dispersive readout has been demonstrated in BLG quantum dots~\cite{banszerus2021dispersive, Reuckriegel_electric2024, Maji2024, ruckriegel2025microwave}, resistive readout remains less explored~\cite{shinozaki2025rfsoc} due to challenges in impedance matching.

Such transport-based sensing requires impedance matching between the resonator including the quantum dot and the 50~$\Omega$ RF circuit.
However, quantum dots in 2D materials typically show large contact resistances~\cite{C7CS00828G, ma2025ohmic}, making impedance matching a major challenge.
Therefore, sensitive resistive readout of BLG devices has been achieved only in field-effect transistor (FET) structures~\cite{johmen2023radio} and quantum point contacts~\cite{hecker2025radio}.

To address this issue, quantum paraelectric varactors based on strontium titanate (SrTiO$_3$) have been proposed as tunable impedance-matching components~\cite{Muller1979STO, Eggli2023Cryo} and have been implemented in RF circuits with carbon-nanotube quantum dots~\cite{apostolidis2024quantum}, which also exhibit large device resistances. 
Because gate-defined 2D quantum dots suffer from similar impedance mismatch issue, extending this approach is a highly promising strategy.

In this study, we integrate an SrTiO$_3$ varactor into an RF reflectometry circuit to enable resistive readout of gate-defined BLG quantum dot and an MoS$_2$ FET device.
We demonstrate nearly perfect impedance matching and observe clear Coulomb oscillations in the reflected RF signal.

\section{II. Results}
\subsection{A. Device structure}

Figure~\ref{fig1}(a) illustrates the stacked structure of the BLG quantum dot device. 
Graphene and hexagonal boron nitride (hBN) flakes are prepared by mechanical exfoliation using a scotch-tape technique, and subsequently stacked onto an undoped silicon substrate by a transfer method employing a polypropylene carbonate–coated polydimethylsiloxane stamp~\cite{pizzocchero2016hot, Iwasaki2020}. 
The bottom graphite layer serves as a back gate, which suppresses stray capacitance from the device structures~\cite{johmen2023radio, shinozaki2025rfsoc}. 
The thicknesses of the top and bottom hBN layers are 14 and 20~nm, respectively. 
The top hBN and BLG layers are etched by reactive ion etching to form edge contacts. 
We employ Ti/Au electrodes as source-drain, side-gate, and finger-gate electrodes deposited by electron-beam evaporation. 
The finger gates are fabricated to be approximately 40~nm wide on a silicon nitride layer deposited by chemical vapor deposition, and are aligned with 40 nm gaps while the gap between the side gate electrodes is also 40 nm.

To form the quantum dot, we define a narrow channel in BLG by applying a vertical electric field between the back and side gates~\cite{zhang2009direct}. 
A Coulomb oscillation is observed by sweeping the finger-gate voltage $V_\mathrm{fg}$ as shown in Fig.~\ref{fig1}(b). 
In this measurement, the back-gate, side-gate, and source–drain voltages are set to $4.4$~V, $-4.5$~V, and $1$~mV, respectively.
All measurements are conducted at a temperature of 2.3 K using a helium decompression refrigerator.
The Coulomb oscillation exhibits peak currents up to $3$~nA, corresponding to a linear conductance of $3~\mathrm{\upmu S}$. 
This value is much smaller than the quantum conductance typically observed in GaAs and ZnO quantum dots~\cite{noro2025charge}. 
Therefore, the conventional circuit design must be modified to satisfy the impedance-matching condition required for high-sensitivity resistive RF detection.

\begin{figure*}
\begin{center}
  \includegraphics{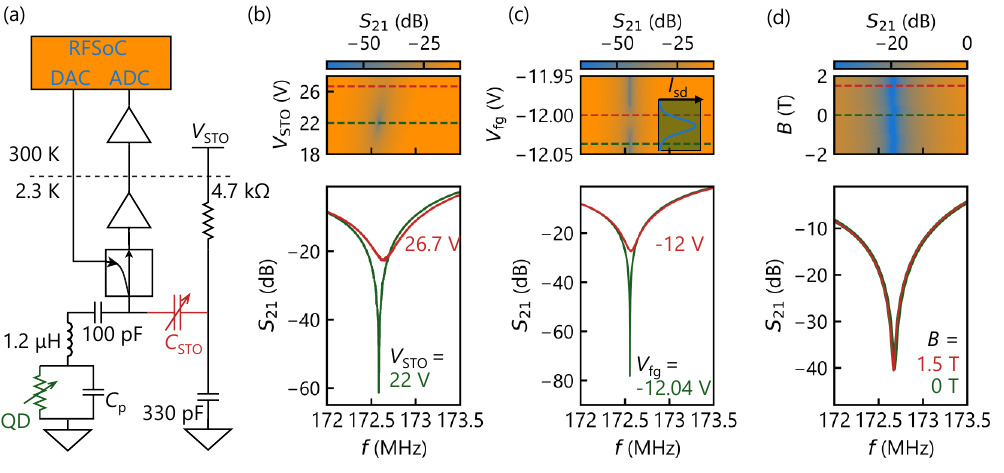}
  \caption{(a) RF reflectometry setup with the RFSoC for impedance matching. 
  (b) Transmission coefficient $S_{21}$ as a function of $V_\mathrm{STO}$, (c) $V_\mathrm{fg}$, and (d) perpendicular magnetic field $B$. 
  The inset in (c) shows a schematic of $I_\mathrm{sd}$ corresponding to the measured $V_\mathrm{fg}$.}
  \label{fig2}
\end{center}
\end{figure*}

\subsection{B. Demonstration of the impedance matching}

The SrTiO$_3$ varactor plays an important role in addressing this issue. 
Figure~\ref{fig2}(a) shows the measurement setup for RF reflectometry. 
The SrTiO$_3$ varactor is connected in parallel with the resonator circuit~\cite{ares2016sensitive, apostolidis2024quantum}, which consists of an inductor, a capacitor, and BLG quantum dots. 
A SrTiO$_{3}$ single crystal with (110) orientation is grown by the Verneuil process (Shinkosha Co.) and annealed for $30$ hours at $1250\,^\circ$C in air. 
The crystal has a thickness of $330~\upmu$m.
At the operating frequency expected to lie between 150 and 200~MHz, the capacitance of the employed varactor is evaluated to range from 40 to 70~pF by varying the bias voltage $V_\mathrm{STO}$, as determined by an on-board calibration technique~\cite{shirachi2025board}.

We utilize radio-frequency system-on-chip (RFSoC) technology to send and receive RF signals, specifically adopting the Zynq\textregistered~UltraScale+\texttrademark~RFSoC ZCU216 evaluation kit~\cite{ZCU216}. 
This RFSoC board is controlled through a Python interface provided by the quantum instrumentation control kit ~\cite{QICK_git, Leamdro_Qick2022, Ding2024exp}, an open-source software framework.
Figure~\ref{fig2}(b) shows the $V_\mathrm{STO}$ dependence of the transmission coefficient $S_{21}$ with $V_\mathrm{fg}$ set near a Coulomb peak. 
The dip in $S_{21}$ is clearly modulated and reaches $-60$~dB at $V_\mathrm{STO}=22$~V. 
With $V_\mathrm{STO}$ tuned to the optimum condition, we then measure the dependence of $S_{21}$ on $V_\mathrm{fg}$, as shown in Fig.~\ref{fig2}(c). 
The range of $V_\mathrm{fg}$ covers a single Coulomb peak, as indicated in the inset of Fig.~\ref{fig2}(c).
The clear change in $S_{21}$ reflects the Coulomb peak and shows that our resonator responds to conductance variations in the BLG quantum dot. 
In particular, the dip approaches $-80$~dB at $V_\mathrm{fg}=-12.04$~V, indicating that the measurement setup achieves nearly perfect impedance matching even though the quantum dot has a large resistance.
A previous study unveiled that the SrTiO$_3$ varactor is insensitive to an in-plane magnetic field, whereas GaAs varactors exhibit a strong dependence on it~\cite{Eggli2023Cryo}. 
In 2D material systems, a perpendicular magnetic field to the device plane introduces the Kramers pairs and is essential for qubit operations.
Therefore, we also confirm the robustness of the resonance characteristics against a perpendicular magnetic field $B$, as shown in Fig.~\ref{fig2}(d). 
The resonance remains robust under $B$, which is consistent with previous findings for the in-plane magnetic field~\cite{Eggli2023Cryo}.

\subsection{C. Demodulation measurement}

\begin{figure}
\begin{center}
  \includegraphics{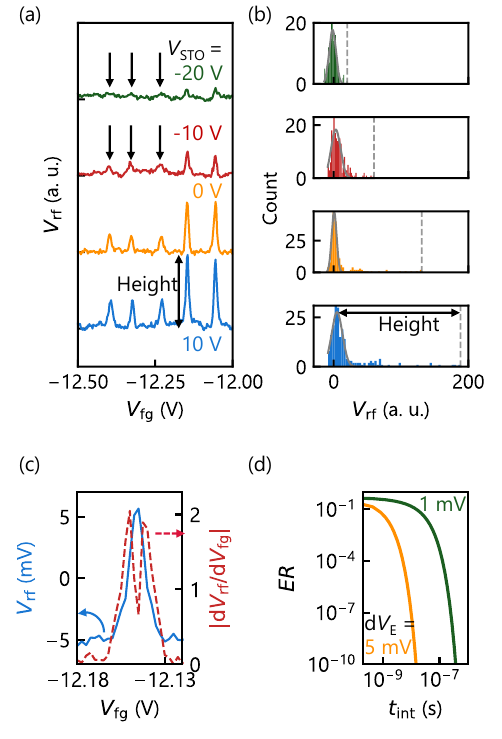}
  \caption{(a) RF-detected Coulomb oscillations for various $V_\mathrm{STO}$. 
  An offset is added to each trace. 
  The black arrows indicate the positions of Coulomb peaks, which are clearly observed at $V_\mathrm{STO}=0$ and $10$~V. 
  (b) Histograms of the Coulomb oscillations. 
  Dashed lines indicate the maximum value of each histogram. 
  Gray traces represent fitted curves obtained using Gaussian functions.
  (c) Close-up view of the Coulomb peak and $\frac{\mathrm{d}V_\mathrm{rf}}{\mathrm{d}V_\mathrm{fg}}$ at $V_\mathrm{STO}=10$~V.  
  (d) Calculated readout error rate for single-electron transitions, evaluated for electrostatic couplings of $\mathrm{d}V_\mathrm{E}=1$~mV and $5$~mV.
}
  \label{fig3}
\end{center}
\end{figure}

Next, we perform demodulation measurements of RF reflectometry. 
An RF signal is applied from the digital-to-analog converter of the RFSoC to the resonator through a directional coupler. 
The reflected signal from the resonator is amplified at both low and room temperatures and digitized by the analog-to-digital converter with a sampling rate of $2.5$~GS/s.
The digitized signal is then demodulated into in-phase and quadrature components within the RFSoC. 
Figure~\ref{fig3}(a) shows Coulomb peaks monitored by the in-phase component, denoted as $V_\mathrm{rf}$, for various $V_\mathrm{STO}$. 
With the system approaching the impedance-matching condition, Coulomb peaks begin to appear.
In particular, smaller peaks become clearly visible near the impedance-matching condition, whereas they vanish far from it.
We convert the RF-detected Coulomb peaks into a histogram as shown in Fig.~\ref{fig3}(b). 
An offset is subtracted to set the histogram at zero.
We define the noise and the peak height as the distribution $\sigma$ extracted from Gaussian fitting and the difference between the maximum and average values of the histogram, respectively.
The peak height at $V_\mathrm{STO}=10$~V is $9.4$ times larger than that at $V_\mathrm{STO}=-20$~V, while their $\sigma$ values remain nearly constant.

Figure~\ref{fig3}(c) shows the replotted $V_\mathrm{rf}$ and $\frac{\mathrm{d}V_\mathrm{rf}}{\mathrm{d}V_\mathrm{fg}}$ at $V_\mathrm{STO}=10$~V.
Under the impedance-matching condition, the RF signal is not reflected from the resonator, resulting in $V_\mathrm{rf}=0$~V. 
As the Coulomb peak crosses $V_\mathrm{rf}=0$~V, the charge detection sensitivity, which is proportional to $\frac{\mathrm{d}V_\mathrm{rf}}{\mathrm{d}V_\mathrm{fg}}$, reaches its maximum value. 
We calculate the potential readout error rate $ER$ assuming a single charge transition as shown in Fig.~\ref{fig3}(d) by using following equation.
\begin{equation}
ER = \frac{1}{2} \mathrm{erfc} \left(\frac{1}{2\sqrt{2}}\frac{\frac{\mathrm{d}V_\mathrm{rf}}{\mathrm{d}V_\mathrm{fg}}\mathrm{d}V_\mathrm{E}}{\sigma_0\sqrt{t_\mathrm{s}/t_\mathrm{int}}}\right),
\label{ErrorRate}
\end{equation}
Here, $\mathrm{d}V_\mathrm{E}$ is the effective gate voltage induced by electrostatic coupling between sensor and target quantum dots, $\sigma_{0}$ the noise distribution without time integration, $t_\mathrm{s}$ the sampling time, and $t_\mathrm{int}$ the integration time for noise reduction.
We assume $\frac{\mathrm{d}V_\mathrm{rf}}{\mathrm{d}V_\mathrm{fg}}=2$ and $\mathrm{d}V_\mathrm{E}$ values of 1~mV and 5~mV for single electron change in the target quantum dots~\cite{kurzmann2019charge}.
These results suggest that impedance matching enables BLG quantum dots to detect single-electron transitions in target quantum dots with high speed.

\subsection{D. Noise response of the varactor}

\begin{figure}
\begin{center}
  \includegraphics{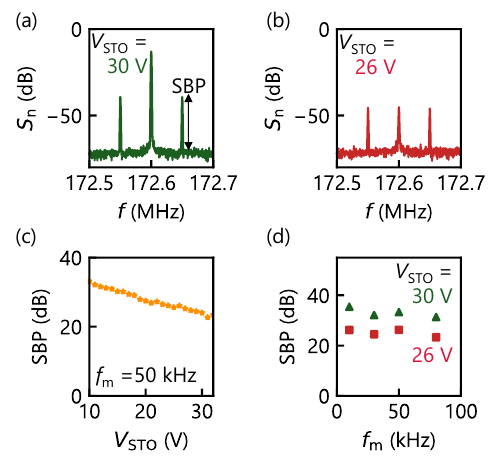}
  \caption{(a) Amplitude-modulation response of the measurement circuit to the SrTiO$_3$ varactor for $V_\mathrm{STO}=30$~V and (b) $26$~V, where the latter corresponds to a condition close to impedance matching. 
  The SBP is defined as the height of the sidebands with respect to the noise background, as indicated. 
  (c) $V_\mathrm{STO}$ dependence of the SBP at $f_\mathrm{m}=50$~kHz. 
  (d) $f_\mathrm{m}$ dependence of the SBP for $V_\mathrm{STO}=30$~V and $26$~V.}
  \label{fig4}
\end{center}
\end{figure}

The ability to tune the impedance matching also makes the system sensitive to the noise of $V_\mathrm{STO}$. 
We apply a sinusoidal signal with a root-mean-square amplitude $V_\mathrm{rms}=1$~V and frequency $f_\mathrm{m}$ in addition to $V_\mathrm{STO}$. 
The power spectrum detected after amplification shows sidebands split by $f_\mathrm{m}$ from the carrier frequency, as shown in Fig.~\ref{fig4}(a) and (b). 
We also define the $\mathrm{SBP}$ as the height of the sidebands above the noise floor. 
The carrier signal at $V_\mathrm{STO}=26$~V is smaller than that at $V_\mathrm{STO}=30$~V, reflecting the low reflection cofficient of the resonator with impedance matching. 
Under the matching condition, the sidebands appears even when the carrier signal is small. 
We summarize the $V_\mathrm{STO}$ dependence of the $\mathrm{SBP}$ in Fig.~\ref{fig3}. 
It decreases monotonically with increasing $V_\mathrm{STO}$ and is proportional to $\frac{\mathrm{d}C_\mathrm{STO}}{\mathrm{d}V_\mathrm{STO}}$ of the SrTiO$_3$ varactor, reflecting the nonlinearity of its dielectric constant~\cite{shirachi2025board}.
Figure~\ref{fig4}(d) indicates the $f_\mathrm{m}$ dependence of $\mathrm{SBP}$ with $V_\mathrm{STO}=26$~V and $V_\mathrm{STO}=30$~V.
For a modulating varactor that tunes the resonance frequency, the $\mathrm{SBP}$ decreases with increasing $f_\mathrm{m}$, indicating the measurement bandwidth for dispersive readout. 
In this case, the $\mathrm{SBP}$ remains almost constant below $100$~kHz, suggesting that the frequency components of the $C_\mathrm{STO}$ noise are directly mapped onto $V_\mathrm{rf}$ at this frequency range.
Here, the capacitance noise $S_\mathrm{C}$ must exceed $1.4~\mathrm{mF}/\sqrt{\mathrm{Hz}}$ to influence $V_\mathrm{rf}$. 
This value is obtained using the following equation~\cite{ares2016sensitive, apostolidis2024quantum},

\begin{equation}
S_{\rm C} = \frac{\Delta C_{\rm STO}}{10^{\rm{SBP}/20} \sqrt{2\Delta f}},
\label{eq1}
\end{equation}
where $\Delta C_\mathrm{STO}$ is the capacitance fluctuation induced by noise and $\Delta f$ the resolution bandwidth.
This value is large compared with that of the varactor used for resonance-frequency tuning~\cite{ares2016sensitive} and indicates robustness to noise contributions from the SrTiO$_3$ varactor to $V_\mathrm{rf}$ even near the impedance-matching condition.

\subsection{E. RF reflectometry with impedance matching for MoS$_2$ devices}
\begin{figure}
\begin{center}
  \includegraphics{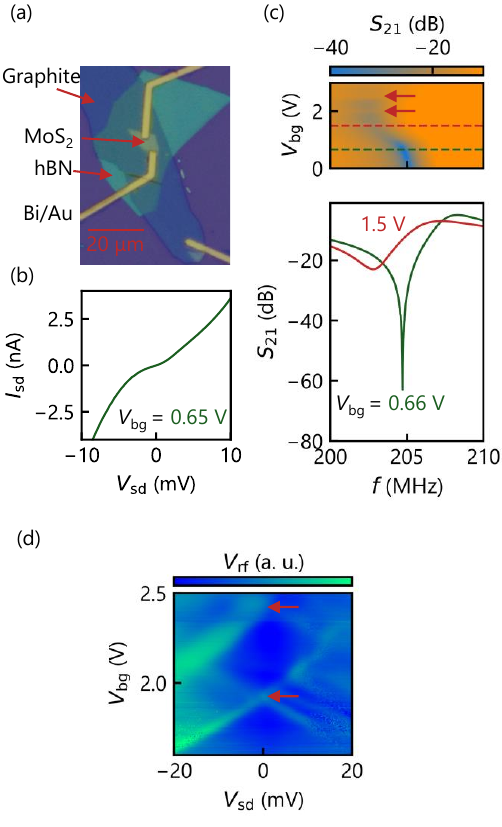}
  \caption{(a) Optical microscope image of the MoS$2$ device. 
  (b) $I\mathrm{sd}$–$V_\mathrm{bg}$ characteristics at $V_\mathrm{sd}=0.65$~V. 
  (c) $V_\mathrm{bg}$ dependence of $S_{21}$. The orange arrows indicate the vanishing points of the resonance. 
  (d) RF-detected Coulomb diamond. 
  The orange arrows correspond to those in (c), indicating the resonance vanishing points.}
  \label{fig5}
\end{center}
\end{figure}

We apply the present technique to a MoS$_2$ device which is also well known to exhibit high resistance at cryogenic temperatures~\cite{kim2017fermi}. 
Figure~\ref{fig5}(a) shows an optical microscope image of the MoS$_2$ FET sturcture.
We stack the graphite/hBN(15~nm)/MoS$_2$(15~nm) layers on the undoped silicon substrate.
The graphite is utilized for the back-gate electrode, as in the BLG devices.
We employ Bi/Au electrodes for the source–drain contacts to obtain near-Ohmic behavior at low bias voltages and cryogenic temperatures~\cite{shen2021ultralow, schock2023non, tataka2024surface, schock2025material}.
This device has linear resistance of $2.8$~M$\Omega$ at $V_\mathrm{bg}$=$0.65$~V, as indicated in Fig.~\ref{fig5}(b) at temperature of 2.3~K.
By setting the optimum $V_\mathrm{STO}$ at $-8.7$~V, we observe the impedance-matching condition in the $V_\mathrm{bg}$ dependence of the $S_{21}$ characteristics, as shown in Fig.~\ref{fig5}(c). 
The resonance shifts with changing $V_\mathrm{bg}$ and vanishes at $V_\mathrm{bg}=1.92$~V and $2.44$~V. 
This shift arises from induced carrier providing additional device capacitance.
Such a quantum dot is unexpected in the simple device structure used in this study. 
A bubble structure in the device and/or local impurity contributions may be considered possible scenarios for the formation of the quantum dot~\cite{johmen2023radio}.
We measure $V_\mathrm{rf}$ as a function of the source–drain voltage $V_\mathrm{sd}$ and the back-gate voltage $V_\mathrm{bg}$. 
A Coulomb diamond with excited states is clearly detected in $V_\mathrm{rf}$ as shown in Fig.~\ref{fig5}(d), and exhibits closing points at $V_\mathrm{bg}=1.92$~V and $2.44$~V, which correspond to the vanishing points in Fig.~\ref{fig5}(c).

\section{IV. Conclusion}

We have demonstrated high-sensitivity RF reflectometry with impedance matching for gate-defined quantum dots based on two-dimensional materials. 
By incorporating a tunable SrTiO$_3$ varactor into the resonator circuit, nearly perfect impedance matching was achieved even for devices with large resistances on the order of megaohms. 
In bilayer graphene quantum dots, this approach enabled clear detection of Coulomb oscillations in the demodulated signals, possessing high sensitivity for single-electron charge sensing. 
Noise analysis revealed that the noise response is governed by $\frac{\mathrm{d}C_\mathrm{STO}}{\mathrm{d}V_\mathrm{STO}}$ reflecting the dielectric nonlinearity of the SrTiO$_3$ varactor.
The measurement system maintained robustness against the perpendicular magnetic field $B$ and noise on $V_\mathrm{STO}$. 
Furthermore, we applied this technique to MoS$_2$ devices and observed impedance-matching conditions as well as RF-detected Coulomb-diamond originating from quantum-dot formation.
These results establish that SrTiO$_3$-based varactors promise impedance-matched RF reflectometry in gate-defined quantum devices based on two-dimensional materials, paving the way for high-speed readout of spin and valley qubits.

\section{Acknowledgements}

We thank Y. Maeda, RIEC Fundamental Technology Center and the Laboratory for Nanoelectronics and Spintronics for technical support. 
Part of this work is supported by 
Grants-in-Aid for Scientific Research (23K26482, 23H04490, 25H01504, 25H02106), 
JST FOREST (JPMJFR246L),
JST CREST (JPMJCR23A2),
Mitsubishi Foundation Research Grant,
Yasumi Science and Technology Foundation Research Grant, 
HABATAKU Young Researchers Support Program,
TIA “KAKEHASHI” program,
and FRiD Tohoku University.
MANA and AIMR are supported by World Premier International Research Center Initiative (WPI), MEXT, Japan.

\section*{Data Availability Statement}
The data that support the findings of this study are available from the corresponding authors upon reasonable request.

\bibliography{reference.bib}

\end{document}